# Accuracy of Real-Time Echo-Planar Imaging Phase Contrast MRI

Pan Liu[1], Sidy Fall[2], Olivier Baledent[2,3]

[1]University of Picardie Jules Verne/CHIMERE EA 7516, Amiens, France
[2]Facing Faces Institute/CHIMERE EA 7516, Amiens, France
[3]Medical image processing department, University hospital Amiens, Amiens, France

## Synopsis

Compared with CINE phase contrast MRI (CINE-PC), echo-planar imaging phase contrast (EPI-PC) can achieve real-time quantification of blood flow, with lower SNR. In this study, the pulsating real model of the simulated cerebral vasculature was used to verify the accuracy of EPI-PC. The imaging time of EPI-PC was 62ms/image at 100*60 spatial resolution. The reconstructed EPI-PC flow curve was extracted by homemade post-processing software. After comparison with the CINE-PC flow curve, it was concluded that EPI-PC can provide an average flow with less than 3% error, and its flow curve will be similar to the CINE-PC flow curve in shape.

## Introduction

Today, CINE phase-contrast MRI (CINE-PC) [1] is the primary noninvasive method for quantifying blood flow and cerebrospinal fluid (CSF) in the cerebrovascular system by using cardiac synchronization [2]. However, it cannot provide the effect of respiration on blood/LCS flow [3] as it can only provide the average cardiac cycle reconstructed using all cardiac cycles obtained during the acquisition. Echo-planar imaging phase contrast (EPI-PC) [4] is a recent technique can achieve real-time quantification of blood/LCS flow and does not require synchronization. Meanwhile, EPI-PC is sensitive to noise and has a low signal-to-noise ratio (SNR). The purpose of this study was to validate the accuracy of EPI-PC in vitro and to quantify the effect of pixel size on both sequences.

## Material

The study was performed on a clinical 3T scanner (Philips Achieva, maximum gradient 80mT/m, gradient increase rate 120mT/m/ms) with a 32-channel head coil.

A flow phantom consisting of 4 tubes in series with different diameters ($\Phi$:9.5/6.4/4.4/2 mm) was attached to the pump to mimicking the vessels in craniospinal system (Figure.1), and a static water-filled tube was placed next to the tubes for calibration. The water pump provides pulsatile flow at 99 bpm with an average flow rate of 1150 mm$^3$/s. This experiment will image the first tube ($\Phi$9.5 mm) with CINE-PC and EPI-PC. CINE-PC provides 32 phase images to show an average pulse cycle. And EPI-PC contains 150 phase images to show the flow rate in real time (Figure 2).

Post-processing is performed by home-made software [5] to extract the flow curves. Calibration is performed by measuring the average velocity of the static tube. In addition, for comparison with the CINE-PC, The EPI-PC flow signal is reconstructed as flow curve with 32 sampling points in the same format as the CINE-PC flow curve.

## METHODS

Both sequences were imaged 10 times using the default parameters (Figure 3) for tube-1. The average flow rate of the pump (1150 mm$^3$/s) and the cross-sectional area of the tube (70.8 mm$^2$) were used as gold standards, and the confidence intervals for the average flow rate and the segmentation area were set to $100 \pm 10\%$ (average flow: [1035 mm$^3$/s: 1265 mm$^3$/s]; segmentation area: [63.72 mm$^2$: 77.8 mm$^2$]). The Bland & Altman test was then used to verify the similarity of the shapes of these two curves.

Based on the default parameters, pixel size was used as a variable for 10 group images (from 0.8 mm to 4.4 mm with 0.4 mm interval), and each group was repeated four times to observe its effect on the two sequences.

## Results

After 10 measurements, reconstructed EPI-PC flow curves were synchronized with CINE-PC flow curves at the peak flow, as shown in Figure 4 A. The average flow of EPI-PC was $1116 \pm 24.5$ mm$^3$/s with a coefficient of variation of 2.3%. The average flow of CINE-PC was $1239 \pm 26.3$ mm3/s with a coefficient of variation of 2.3%. Both the average flow and the segmentation area (EPI-PC is 70.1 mm$^2$ and CINE-PC is 70 mm$^2$) are within the confidence interval. The Bland-Altman plot (Figure 4 B) reflects that the differences of the 32 points corresponding to the two curves are within the limits of agreement (LoA), which can indicate that the two curves have similar shapes.

Figure 5 shows that the average flow of EPI-PC is within the confidence interval when the pixel size is between 1.2 mm and 2.8 mm, while the pixel size of CINE-PC needs to be set between 1.2 mm and 2.4 mm to provide an accurate average flow.

## Discussion

Under the present experimental conditions, the reconstructed EPI-PC flow curve can replace the CINE-PC flow curve. However, in vivo, cerebral arterial blood flow waves usually have higher frequency components (4Hz-5Hz), and for the current sampling frequency, EPI-PC may have difficulty in correctly representing these high frequency waveforms.

CINE-PC completes multiple phase encodings in each pulse cycle and fills them in K-space of the corresponding phase images, so that even if the acquisition time increases, it does not affect its pseudo-sampling interval Δt. However, for EPI-PC with real-time imaging, Δt is directly related to the acquisition time. Therefore, increasing the sampling frequency can improve the accuracy of the reconstructed curve shape without affecting the accuracy of the results. In the case of larger pixel size, not only the signal-to-noise ratio can be improved, but also the sampling frequency can be increased. However, if the pixel size is too large, a partial volume effect will occur, which distorts the results. The findings of this study indicate that setting the pixel size of EPI-PC between 25%-30% of the target vessel/LCS diameter can obtain a higher signal-to-noise ratio and a higher temporal resolution.

## Conclusion

This experiment demonstrates the feasibility of using EPI-PC for in vivo real-time imaging. After post-processing with specific software, EPI-PC can provide the same flow curve as CINE-PC and obtain time-dimensional data. Increasing the sampling frequency allows for better display of the details of the reconstructed flow curve from EPI-PC. With the application of more acceleration techniques such as compressed sensing, EPI sequences will have greater potential for clinical diagnosis and for studying the effect of respiration on blood flow.

## Acknowledgements

This research was supported by EquipEX FIGURES (Facing Faces Institute Guilding Research), European Union Interreg REVERT Project, Hanuman ANR-18-CE45-0014 and Region Haut de France.

Thanks to the staff members at the Facing Faces Institute (Amiens, France) for technical assistance.

Thanks to David Chechin from Phillips industry for his scientific support.

## References

[1] Nj, Pelc, Herfkens Rj, Shimakawa A, and Enzmann Dr. Phase Contrast Cine Magnetic Resonance Imaging. Magnetic Resonance Quarterly 7, no 4 (1 octobre 1991): 229 54.

[2] Balédent, O., Henry-Feugeas, M. C. and Idy-Peretti, I. Cerebrospinal fluid dynamics and relation with blood flow: a magnetic resonance study with semiautomated cerebrospinal fluid segmentation. Invest Radiol 36, (2001): 368–377.

[3] Chen, Liyong, Alexander Beckett, Ajay Verma, and David A. Feinberg. Dynamics of Respiratory and Cardiac CSF Motion Revealed with Real-Time Simultaneous Multi-Slice EPI Velocity Phase Contrast Imaging. NeuroImage 122 (15 novembre 2015): 281 87.

[4] Pan LIU, Armelle LOKOSSOU, Sidy FALL, Malek MAKKI, and Olivier BALEDENT. Post Processing Software for Echo Planar Imaging Phase Contrast Sequence. ISMRM 27th, (2019): N. 4823.

[5] Tang Chao, Duane D. Blatter, et Dennis L. Parker. Accuracy of Phase-Contrast Flow Measurements in the Presence of Partial-Volume Effects. Journal of Magnetic Resonance Imaging 3, (1993): no 2.

## Figures

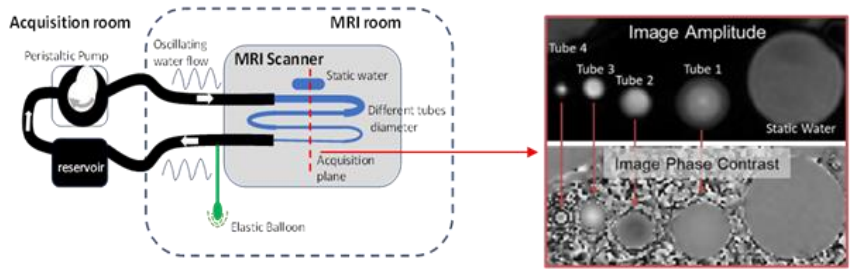

Figure 1: **Left**: A pulsatile flow phantom simulating the vasculature of the cranial system is located in the center of the head coil. An elastic balloon is connected in parallel to the outflow tube for connection to the synchronizer of the CINE-PC. **Right**: Amplitude and phase images of the 4 tubes and the static water-filled tube in the imaging plane.

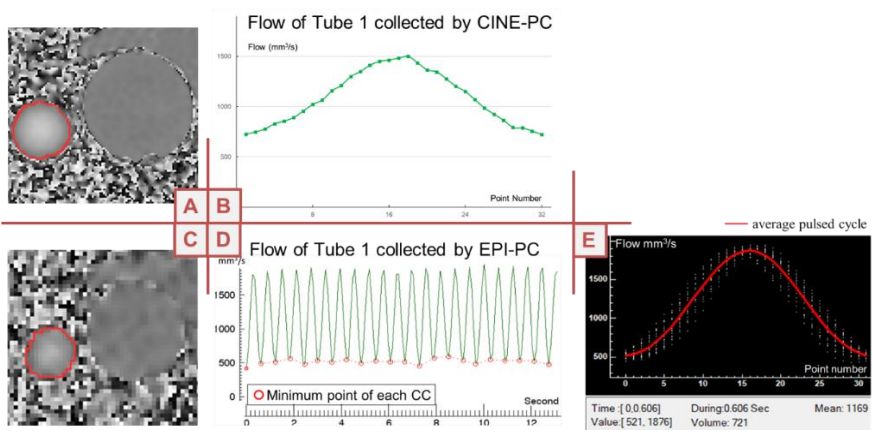

Figure 2: CINE-PC phase image A) and its calibrated flow curve B). EPI-PC phase image C) and its calibrated flow curve D) with the minimum points (red cycles) in each cycle automatically found by the software which will be used as the 'cutting' position, all the pulse cycles of EPI-PC will be used to reconstruct an average pulsed cycle E).

| | CINE-PC | EPI-PC |
|---|---|---|
| FOV(F*P) | 100*60 | 100*60 |
| Venc (mm/s) | 50 | 50 |
| Pixel size (mm x mm) | 1.2x1.2 | 1.2x1.2 |
| Thickness (mm) | 4 | 4 |
| Flip angle (degrees) | 30 | 30 |
| EPI-factor | -- | 9 |
| SENSE (P-factor) | 1.5 | 2.5 |
| TR (ms) | 11 | 15.2 |
| TE (ms) | 7.7 | 9.1 |
| Acquisition time (s) | 23.6 | 9.3 |
| Nb images / Cycle | 32 | 9.7 |

Figure 3: Default parameters of CINE-PC and EPI-PC

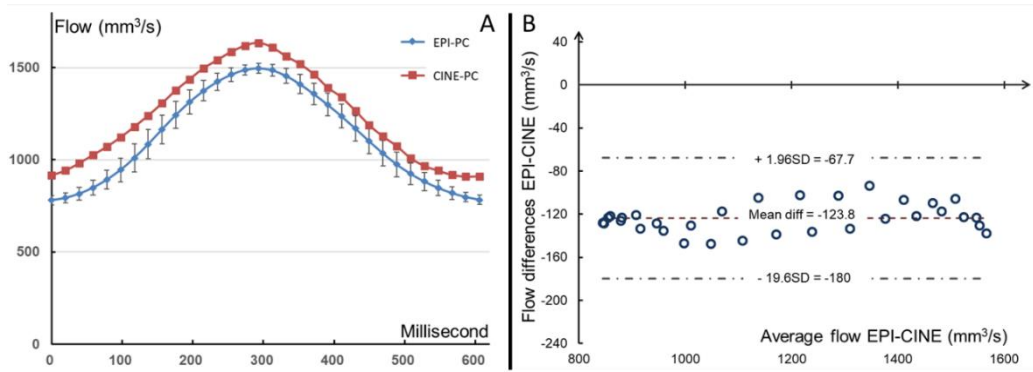

Figure 4: Average flow curve of CINE-PC and of EPI-PC. The error bar in the EPI-PC curve represents the standard deviation of each corresponding point. B: Bland–Altman plot graphic of EPI-PC curve and CINE-PC curve. The X axis represents the average flow of the corresponding points in the two sequences and the Y axis represents the difference between the corresponding points of two sequences. The red line represents the average difference of the 32 points which is -123.8 mm³/s, 95% limits of agreement (LoA) calculated by average of flow differences ± 1.96 × standard deviation of the differences, that is -67.7 mm³/s and 180 mm³/s.

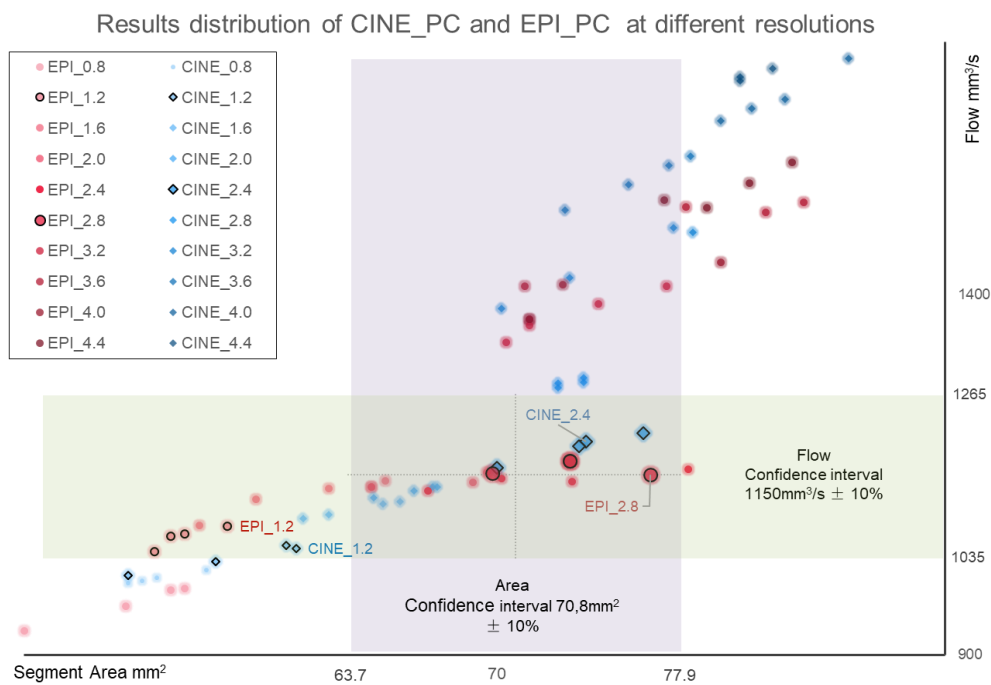

Figure 5: Results (segmentation area & average flow) distribution of EPI-PC (red points) and CINE-PC (blue points) under different pixel size (different color saturation). The X axis represents the segmentation area, the purple shade is the confidence interval of segmentation area. The Y axis represent the flow rate, the green shade is the confidence interval of flow rate.